\documentstyle[a4,12pt]{article}
\begin{document}
\hoffset 0.25in
\setcounter{page}{1}
\begin{flushright}
LPTENS 93/14 \\ ROME prep. 93/946\\ LPTHE Orsay-93/20\\ SHEP 92/93-25\\ WUB
93-27 July 1993.
\end{flushright}

\newcommand{\be}{\begin{equation}}
\newcommand{\ee}{\end{equation}}
\newcommand{\bea}{\begin{eqnarray}}
\newcommand{\eea}{\end{eqnarray}}
\newcommand{\nn}{\nonumber}
\newcommand{\muh}{\hat\mu}
\newcommand{\dlr}{\stackrel{\leftrightarrow}{D} _\mu}
\newcommand{\vnew}{$V^{\rm{NEW}}$}
\newcommand{\vecp}{$\vec p$}
\newcommand{\dof}{{\rm d.o.f.}}
\newcommand{\prd}{Phys.Rev. \underline}
\newcommand{\pl}{Phys.Lett. \underline}
\newcommand{\prl}{Phys.Rev.Lett. \underline}
\newcommand{\np}{Nucl.Phys. \underline}
\newcommand{\vvp}{v_B\cdot v_D}
\newcommand{\dl}{\stackrel{\leftarrow}{D}}
\newcommand{\dr}{\stackrel{\rightarrow}{D}}
\newcommand{\mev}{{\rm MeV}}
\newcommand{\gev}{{\rm GeV}}
\newcommand{\calp}{{\cal P}}
\pagestyle{empty}
\centerline{\LARGE{\bf{Semi-leptonic Decays of Heavy Flavours}}}
\centerline{\LARGE{\bf{ on a Fine Grained Lattice}}}
\vskip 1.5cm
\centerline{\bf{As.Abada$^1$, C.R.Allton$^2$, Ph.Boucaud$^1$,
D.B.Carpenter$^3$,}}
\centerline{\bf{M.Crisafulli$^2$, S.G\"usken$^4$,
P.Hernandez$^5$, V.Lubicz$^2$, G.Martinelli$^{2,6}$,}}
\centerline{\bf{O.P\`ene$^1$, C.T.Sachrajda$^7$, K.Schilling$^4$,
G.Siegert$^4$, R.Sommer$^8$.}}
\vskip 0.5cm
\centerline{$^1$ LPTHE, Orsay, France,\footnote {Laboratoire associ\'e au
 Centre National de la Recherche Scientifique.}}
\centerline{$^2$ Dip. di Fisica, Univ. di Roma \lq La Sapienza\rq,}
\centerline{I-00185 Roma,Italy and INFN, Sezione di Roma, Italy.}
\centerline{$^3$ Dept. of Electronics and Computer Science,}
\centerline{The University, Southampton SO9 5NH, UK.}
\centerline{$^4$ Physics Dept., Univ. of Wuppertal,
D-42097 Wuppertal, Germany. }
\centerline{$^5$ Dept. de F\'is. Te\'orica C-XI,
Univ. Aut\'onoma de Madrid, E-28049, Madrid ,Spain.}
\centerline{$^6$ Laboratoire de Physique Th\'eorique de
l'\'Ecole Normale Sup\'erieure\footnote {Unit\'e Propre du Centre National de
la Recherche Scientifique, associ\'ee \`a l'\'Ecole Normale Sup\'erieure et
\`a l'Universit\'e de Paris-Sud.}.}
\centerline{24 rue Lhomond, 75231 Paris CEDEX 05, France.}
\centerline{$^7$ Dept. of Physics, The University, Southampton SO9 5NH, UK.}
\centerline{$^8$ Deutsches Elektronen-Synchrotron, DESY,
Notkestrasse 85,D-22603 Hamburg, Germany.}


\begin{abstract}
We present the results of a numerical calculation of semi-leptonic form
factors relevant for heavy flavour meson decays into light mesons.  The
results have been obtained by studying two- and three-point correlation
functions at $\beta=6.4$ on a $24^3 \times 60$ lattice, using the Wilson
action in the quenched approximation. \\ {}From the study of the matrix element
$<K^-\vert J_\mu \vert D^0>$ we obtain $f^+_K(0)=0.65\pm 0.18$. From the
matrix element $<\bar K^{* 0}\vert J_\mu \vert D^+>$ we obtain $V(0)=0.95\pm
0.34$, $A_1(0)=0.63\pm 0.14 $ and $A_2(0)=0.45\pm 0.33 $. We also obtain
$A_1(q^2_{max})=0.62\pm 0.09$, $V(0)/A_1(0)=1.5\pm 0.28 $ and
$A_2(0)/A_1(0)=0.7\pm 0.4$. The results for $f^+_K(0)$, $V(0)$ and $A_1(0)$
are consistent with the experimental data and with previous lattice
determinations with larger lattice spacings. In the case of $A_2(0)$ the
errors are too large to draw any firm conclusion.  We also show, with the help
of
the heavy quark effective theory (HQET), that it is
possible to extrapolate the form factors to the B meson.
 Our calculations show that the form factors
follow a behaviour compatible with the predictions from HQET.
Within large uncertainties, our results suggest
that $A_2/A_1$ increases with the heavy quark mass. For B mesons $A_2/A_1$
can be as large as $1.5$-$2.0$. We also get very rough estimates for the
partial decay widths $\Gamma (B \rightarrow \pi l \nu_l)=\vert V_{ub} \vert^2
(12 \pm 8) \times 10^{12} s^{-1}$ and $\Gamma(B \rightarrow \rho l
\nu_l)=\vert V_{ub} \vert^2 (13 \pm12) \times 10^{12} s^{-1} $, which can be
used to give upper bounds on the rates.

\end{abstract}
\newpage
\pagestyle{plain}
\section{Introduction}
\label{sec:intro}
Semi-leptonic decays of heavy-light mesons have attracted considerable
interest in the past years as they play a crucial role in the determination
of the Cabibbo-Kobayashi-Maskawa mixing matrix and in the understanding of
weak decays. Moreover, the study of the dependence of the form factors on the
heavy quark mass checks the validity of the scaling laws, predicted by the
heavy quark effective theory, in the range of masses corresponding to D and B
mesons.

There is increasing evidence, that lattice QCD allows {\it ab initio}
quantitative predictions of weak decay matrix elements\cite{review}.
Exclusive semi-leptonic decay channels of heavy flavour mesons have been
studied in a series of papers\cite{victor}-\cite{bes2}, at a lattice
resolution of $a^{-1} \sim 2-3$ GeV.  In order to improve the control over
discretization errors, we present here a study on pseudoscalar-pseudoscalar
as well as pseudoscalar-vector weak form factors, on a lattice $24^3 \times
60$, at $ \beta = 6.4$, which corresponds to $a^{-1}\simeq 3.6-3.7$ GeV.
This is a continuation of our recent work\cite{orsw} on the leptonic decay
constants.

The main results of the present study are given in the abstract and in tables
\ref{tab:final}, \ref{tab:largeurs} and \ref{tab:extrab}.  These results have
been obtained by extrapolating the matrix elements in momentum transfer, in
the light and heavy quark masses. We have included in tables
\ref{tab:emrps}-\ref{tab:semifinal} all the numbers obtained from the
direct study of the two- and three- point correlation functions, computed at
several values of $K_W$.  These tables may be useful to check our results. We
also give the details of the extrapolation, in order to allow the reader to
reproduce the form factors in table
\ref{tab:final} from the meson masses and the form factors of tables
\ref{tab:emrps}-\ref{tab:semifinal}. In section
\ref{sec:slb} we show that it is possible to extrapolate the form factors
 to the B meson using the Heavy Quark Effective Theory (HQET). In this way
one can get also informations on the corrections to the infinite mass limit.
Due to large statistical errors, we are only able to give a rough estimate
for the branching ratios of the processes $B \rightarrow \pi$ and $B
\rightarrow \rho$.
\section{Description of the Calculation}
\label{sec:des}
\subsection{Lattice Setup}
We work with the standard Wilson action at $\beta=6.4$ for the gauge fields
and the quark propagators\cite{wi}, in the quenched approximation.  We have
generated 15 independent gauge field configurations on a $24^3 \times 30$
lattice, separated by $500$-$1600$ sweeps, using the overrelaxed
algorithm\cite{cr}. The 15 configurations were produced in groups of 5, with
three independent initial conditions. In each of the three cases the first of
the useful configurations was obtained after an initial thermalization of at
least $3000$ sweeps ( $500$ with the Metropolis algorithm and $2500$ with the
overrelaxed algorithm ).
\par On each configuration we have computed the quark propagators for 7
different values of the Wilson hopping parameter $K_W$, corresponding to
\lq\lq heavy" quarks, $K_H=0.1275$, $0.1325$, $0.1375$, $0.1425$,
and \lq\lq light" quarks, $K_L=0.1485$, $0.1490$ and $0.1495$.  Periodic
boundary conditions on a $24^3 \times 60$ lattice have been imposed in the
calculation of the quark propagators, by using appropriate combinations of
periodic and antiperiodic (in the time direction) quark propagators
calculated on a $24^3 \times 30$ lattice\cite{sc}.  Further details on the
lattice calibration, fitting procedures, mass spectrum, extraction of matrix
elements of local operators between the vacuum and meson states, e.g.
$<M_P\vert \bar Q \gamma_5 q\vert 0>$, can be found in ref.\cite {orsw}.
\subsection{Form Factor Evaluation}
\par
{}From the study of three-point correlation functions
\cite{victor}-\cite{bes2}, one extracts the weak current matrix elements
for a given momentum transfer:
\be < K \vert J_\mu \vert D > = \Bigl( p_D + p_K -
\frac{ M_D^2 - M_K^2}{q^2} q \Bigr)_\mu f^+_K(q^2) +
\frac{ M_D^2-M_K^2}{q^2} q_\mu f^0_K(q^2) \label{dk}\ee
\bea
 < K^*_r \vert J_\mu \vert D > &=& e^\beta_r \Bigl[ \frac{2
V(q^2)}{M_D+M_{K^*}}
\epsilon_{\mu\gamma\delta\beta}p_D^\gamma p_{K^*}^\delta + i
( M_D +M_{K^*}) A_1(q^2) g_{\mu\beta} \nn \\ &-& i
\frac{A_2(q^2)}{M_D+M_{K^*}}P_\mu q_\beta + i \frac{A(q^2)}{q^2}2
M_{K^*}q_\mu P_\beta \Bigr] \label{dks} , \eea where $q$ is the momentum
transfer, $q=p_D-p_K$ or $q=p_D-p_{K^*}$, $P=p_D+p_{K^*}$ and $e^\beta_r$ is
the polarization vector of the $K^*$.  $f^{+,0}_K$, $V$, $A_{1,2}$ and $A$
are dimensionless form factors in the helicity basis, see for example
\cite{lubicz}.  {}From the matrix elements (\ref{dk}) and (\ref{dks}), by
varying the Lorentz component of the current, the meson momenta and the $K^*$
polarization, one can extract the form factors.

In the following we give those details of the calculation which are specific
to semi-leptonic decays and cannot be found in ref.\cite{orsw}.
\par The matrix elements
have been computed for a D meson at rest using a pseudoscalar density as
source, at a time distance $(t_D-t_{K,K^*})/a=28$.  The position of the light
meson source is fixed in the origin and we have varied the time position (in
the interval $t_J/a =12-16$) and the momentum of the weak current. For the
$K$-source we have also used the pseudoscalar density and for the $K^*$ the
local vector current. For the weak axial current we have chosen the local
operator $Z_A \bar Q(x) \gamma_\mu \gamma_5 q(x)$, while for the weak vector
current we have used both the local ($Z_V \bar Q(x) \gamma_\mu q(x)$) and the
``conserved " currents\footnote{ The latter would be conserved on the lattice
in the limit of degenerate quark masses\cite{ks,boc}.}. $Z_A$ and $Z_V$ are
the renormalization constants of the axial and vector current
respectively\cite{ks,boc}.  The values quoted for the form factors are based
on the choice $Z_A=0.88$ and $Z_V=0.84$\cite{mz}-\cite{smit} which will be
justified below.
\par The two- and three-point  correlation functions have been computed
for different spatial momenta which are allowed by the lattice discretization
and volume. By defining $\vec p_{K,K^*}=2 \pi / (La) \times ( n_x, n_y
,n_z)$, with $L=24$, we have used for $\vec p_{K,K^*}$ the following
assignments $(0,0,0)$, $(1,0,0)$, $(0,1,0)$, $(0,0,1)$, $(1,1,0)$, $(1,0,1)$,
..., $(2,0,0)$, ..., and $(1,1,1)$. We have combined different correlation
functions which can be related by the spatial orthogonal group and parity.
In all we have five independent momenta which in the following we will denote
by $(0,0,0)$, $(1,0,0)$, $(2,0,0)$, $(1,1,0)$ and $(1,1,1)$.

\par Large momenta in lattice units imply large systematic errors.
 In order to monitor lattice artefacts at large momenta, we have fitted the
light pseudoscalar meson two-point functions to the asymptotic form (at large
time distances):

\bea G_5(t,\vec p) &=& \sum_{\vec x} e^{i\vec p \cdot\vec x} < P_5(\vec x, t)
P_5^{\dagger}(\vec 0,0)> \nn \\ &\rightarrow& Z_{5} \frac{e^{-E_5 T /2}}{E_5}
\cosh\Bigl( E_5 (t-T/2)\Bigr) , \label{fit} \eea
 where $P_5= \bar q(x)
\gamma_5 q^{\prime}(x)$, $T$ is the time lattice length  and $E_5$ the meson
energy. The corresponding formula for the
vector meson can be easily derived:
\bea G_{ij}(t,\vec p) = \sum_{\vec x} e^{i\vec p \cdot\vec x} < V_i(\vec x,
t) V_j^{\dagger}(\vec 0,0)> \rightarrow \nn \\ \Bigl( - g_{ij} + \frac{p_i
p_j} {M_V^2} \Bigr) Z_{VV} \frac{e^{-E_V T /2}}{E_V}
\cosh\Bigl( E_V (t-T/2)\Bigr) , \label{fitv} \eea
 where $V_i= \bar q(x) \gamma_i q^{\prime}(x)$.
\begin{table}
\centering
\begin{tabular} {|c|c|c|c|c|}
\hline
{\em $K_1 = K_2$ } &momentum& $E_5$ & $\phantom {\bigg(}\bar E_5\phantom
{\bigg(}$ & $Z_{5} \times 10^3$\\
\hline
{\em} & $0,0,0$& 0.27(1) & 0.27(1) & 3.9(0.3) \\ 0.1485 & $1,0,0$& 0.39(2) &
0.38(1) & 5.7(1.6) \\ {\em} & $1,1,0$& 0.41(4) & 0.46(1) & 2.5(1.6) \\ \hline
{\em} & $0,0,0$& 0.23(1) & 0.23(1) & 3.3(0.3) \\ 0.1490 & $1,0,0$& 0.37(3) &
0.35(1) & 5.4(2.5) \\ {\em} & $1,1,0$& 0.37(5) & 0.44(1) & 1.7(1.3) \\ \hline
{\em} & $0,0,0$& 0.19(1) & 0.19(1) & 2.9(0.4) \\ 0.1495 & $1,0,0$& 0.36(5) &
0.33(1) & 5.3(5.2) \\ {\em} & $1,1,0$& 0.32(5) & 0.42(1) & 1.1(0.9) \\ \hline
\end{tabular}
\caption{\it{Pseudoscalar meson masses and  energies
in lattice units and $Z_{5}$ from a fit to two-point correlations at
different momenta and quark masses.  The mesons are composed of two
degenerate light quarks, and the Wilson parameter is given in the first
column.}}
\label{tab:emrps}
\end{table}

 An estimate of the lattice effects is obtained from a comparison of $E_5$,
obtained from eq.(\ref{fit}), with $\bar E_5=\sqrt{M_5^2+ \vec p^2}$, where
$M_5$ is the value of $E_5$ found from a fit at $\vec p=0$. Moreover the
values of $Z_{5}$ obtained by fitting the two-point functions at different
momenta should be consistent (similarly for vector mesons).  In analyzing the
two-point correlation functions, we found that for $(1,1,1)$ and $(2,0,0)$
the results are too noisy with the present statistics and therefore they have
not been used. In tables
\ref{tab:emrps} and \ref{tab:emrv}
we report, for different light quark masses, $E_5$ ($E_V$) and $Z_{5}$
($Z_{VV}$) as obtained using the fit in eq.(\ref{fit}) (eq.(\ref{fitv})), as
well as $\bar E_5$ ($\bar E_V$). The results reported in tables
\ref{tab:emrps} and \ref{tab:emrv} have been obtained by fitting the
two-point functions in the interval $12 \le t/a \le 20$
\begin{table}
\centering
\begin{tabular} {|c|c|c|c|c|}
\hline
{\em $K_1 = K_2$ } &momentum& $E_V$ & $\phantom {\bigg(}\bar E_V\phantom
{\bigg(}$ & $Z_{VV} \times 10^3$\\
\hline
{\em} & $0,0,0$& 0.33(1) & 0.33(1) & 1.5(2) \\ 0.1485 & $1,0,0$& 0.42(1) &
0.42(1) & 1.6(2) \\ {\em} & $1,1,0$& 0.51(1) & 0.49(1) & 2.1(5) \\ \hline
{\em} & $0,0,0$& 0.30(1) & 0.30(1) & 1.2(1) \\ 0.1490 & $1,0,0$& 0.40(1) &
0.40(1) & 1.3(2) \\ {\em} & $1,1,0$& 0.50(2) & 0.48(1) & 2.0(6) \\ \hline
{\em} & $0,0,0$& 0.27(1) & 0.27(1) & 0.9(1) \\ 0.1495 & $1,0,0$& 0.38(1) &
0.38(1) & 1.1(2) \\ {\em} & $1,1,0$& 0.51(2) & 0.46(1) & 2.3(8) \\ \hline
\end{tabular}
\caption{\it{Vector meson masses and  energies
in lattice units and $Z_{VV}$ from a fit to two-point correlations at
different momenta and quark masses.  The mesons are composed of two
degenerate light quarks, and the Wilson parameter is given in the first
column.}}
\label{tab:emrv}
\end{table}
Tables \ref{tab:emrps} and \ref{tab:emrv} show that for the momenta $(1,0,0)$
and $(1,1,0)$ the continuum energy-momentum relation is satisfied within
large statistical fluctuations. The statistical error increases as one
approaches to the chiral limit and the agreement appears to be better for the
vector meson than for the pseudoscalar one. We believe that this is due to
our limited statistics.

\par In order to extract the current matrix elements from the
three-point functions several approaches may be followed.  We have used two
different methods denoted in the following by ``analytic'' and ``ratio''. To
be specific, we only explain these methods in the case of $<K \vert J_\mu
\vert D>$.  The three-point correlation function is given by:
\bea C_\mu(t_x,t_y)&=& \sum_{\vec x, \vec y} < 0 \vert T \Bigl[
P_5^K(\vec 0, 0 ) J_\mu (\vec x , t_x) P_5^{ \dagger D}( \vec y, t_y)\Bigr]
\vert 0 > e^{i \vec q \cdot \vec x + i \vec p_D \cdot \vec y}
\nn \\ &\rightarrow&  \frac {\sqrt{Z^K_{5} Z^D_{5}}}{4 E^K_5 E^D_5}
 < K (\vec p_D+\vec q)\vert J_\mu \vert D (\vec p_D) > e^{-E^K_5 \vert t_x
\vert -E^D_5 \vert t_y-t_x \vert} \label{3pf} \eea at large time distances.
In eq.(\ref{3pf}), $P_5^K$ ($P_5^{\dagger D}$) is the pseudoscalar density
used to annihilate (create) a K meson (D meson) and $E^{K,D}_5$ the
corresponding energies\footnote{We recall that all our calculations have been
done with $\vec p_D=0$.}. The matrix element can be isolated from the ratio:
\be R= \frac {C_\mu (t_x,t_y)}{G_5^K(t_x, \vec q + \vec p_D)
G_5^D(t_y-t_x,\vec p_D)}
\times\sqrt{Z^K_{5}Z^D_{5}} \rightarrow < K \vert J_\mu \vert D >
\label{r3pf}\ee
where $G_5^{K,D}$ are the analog of $G_5$ defined in eq.(\ref{fit}). \par The
two methods to extract the matrix elements are as follows:
\begin{itemize}
\item   ``analytic":  in the denominator of
eq.(\ref{r3pf}), we use the analytic expression of $G^{K,D}_5$ given in
eq.(\ref{fit}), with $Z_5$ and $\bar E_5$ (computed from the meson mass)
taken from the fit to the two-point function at zero momentum.
\item  ``ratio": we divide the three-point
correlation function by the two-point correlation functions with appropriate
momenta averaged over the same configurations. We then multiply the result by
the factor $\sqrt{Z_5^K Z_5^D}$ where $Z_5^K$ and $Z_5^D$ are obtained from
the fit at zero momentum to the two-point functions corresponding to the D
and K propagators.
\end{itemize}
With large time distances and a high statistics the two methods should agree,
up to $O(a)$ effects. Our limited statistics gives fluctuations in the
energy-momentum relations, in addition to these $O(a)$ effects. Therefore the
two methods may yield slighlty different results and we will take into
account the differences in the evaluation of the final error, see below. We
report in tables \ref{tab:ff1} and \ref{tab:ff2} the form factors for our set
of light and heavy quark masses, from the analytic method and using the
conserved vector current.
\begin{table}
\centering
\begin{tabular} {|c|c|c|c|c|}
\hline
 {\em $K_1 , K_2$ } &$\vec p$& $q^2 a^2$ & $f^+(q^2)$ & $f^0(q^2)$\\ \hline
0.1275 & $0,0,0$& 0.296(4) & $- $& 0.91(8) \\ 0.1485 & $1,0,0$& 0.123(3) &
0.81(8) & 0.74(7) \\ {\em} & $1,1,0$& -0.010(2) & 0.76(17) & 0.77(17) \\
\hline 0.1275 & $0,0,0$& 0.325(5) & $ - $ & 0.90(9) \\ 0.1490 & $1,0,0$&
0.137(3) & 0.82(11) & 0.73(9) \\ {\em} & $1,1,0$& -0.003(3) & 0.82(21) &
0.83(21) \\ \hline 0.1275 & $0,0,0$& 0.361(7) & $- $& 0.91(11) \\ 0.1495 &
$1,0,0$& 0.151(4) & 0.82(17) & 0.71(13) \\ {\em} & $1,1,0$& 0.005(3) &
0.97(26) & 0.97(26) \\ \hline 0.1325 & $0,0,0$& 0.186(3) &$-$ & 0.93(8) \\
0.1485 & $1,0,0$& 0.037(2) & 0.79(7) & 0.76(7) \\ {\em} & $1,1,0$& -0.078(1)
& 0.74(17) & 0.80(16) \\ \hline 0.1325 & $0,0,0$& 0.208(4) & $-$ & 0.92(10)
\\ 0.1490 & $1,0,0$& 0.046(2) & 0.78(10) & 0.74(9) \\ {\em} & $1,1,0$&
-0.074(2) & 0.79(21) & 0.87(21) \\ \hline 0.1325 & $0,0,0$& 0.236(6) & $-$ &
0.93(11) \\ 0.1495 & $1,0,0$& 0.056(3) & 0.78(15) & 0.73(13) \\ {\em} &
$1,1,0$& -0.069(2) & 0.94(26) & 1.03(28) \\ \hline 0.1375 & $0,0,0$& 0.097(2)
& $-$ & 0.95(9) \\ 0.1485 & $1,0,0$& -0.027(1) & 0.75(7) & 0.78(7) \\ {\em} &
$1,1,0$& -0.122(1) & 0.70(17) & 0.83(16) \\ \hline 0.1375 & $0,0,0$& 0.112(3)
& $-$ & 0.95(10) \\ 0.1490 & $1,0,0$& -0.021(1) & 0.73(9) & 0.76(9) \\ {\em}
& $1,1,0$& -0.120(1) & 0.75(21) & 0.92(22) \\ \hline 0.1375 & $0,0,0$&
0.132(4) & $-$ & 0.95(11) \\ 0.1495 & $1,0,0$& -0.015(2) & 0.72(13) &
0.74(14) \\ {\em} & $1,1,0$& -0.118(1) & 0.89(26) & 1.10(31) \\ \hline 0.1425
& $0,0,0$& 0.033(1) & $-$ & 0.97(9) \\ 0.1485 & $1,0,0$& -0.063(-) & 0.69(7)
& 0.79(8) \\ {\em} & $1,1,0$& -0.137(-) & 0.65(17) & 0.88(17) \\ \hline
0.1425 & $0,0,0$& 0.042(1) &$-$ & 0.96(10) \\ 0.1490 & $1,0,0$& -0.061(-) &
0.66(8) & 0.77(10) \\ {\em} & $1,1,0$& -0.137(-) & 0.68(21) & 0.98(25) \\
\hline 0.1425 & $0,0,0$& 0.053(2) & $-$ & 0.96(11) \\ 0.1495 & $1,0,0$&
-0.059(1) & 0.62(11)& 0.75(15) \\ {\em} & $1,1,0$& -0.137(-) & 0.80(27) &
1.16(40) \\ \hline
\end{tabular}
\caption{\it{Pseudoscalar $\rightarrow$ pseudoscalar
form factors for different momenta and quark masses using the ``analytic"
method. The values reported in the table have been obtained using the
``conserved" current. Errors denoted by (-) are smaller than the figures
reported in the table. We also give the squared momentum transfer in lattice
units.}}
\label{tab:ff1}
\end{table}
\begin{table}
\centering
\begin{tabular} {|c|c|c|c|c|c|c|}
\hline
{\em $K_1 , K_2$ } &$\vec p$& $q^2 a^2$ & $V(q^2)$ & $A_1(q^2)$ & $A_2(q^2)$
&$A(q^2)$\\ \hline 0.1275, & $0,0,0$& 0.237(6) & $-$ & 0.61(6) & $-$& $-$ \\
0.1485 & $1,0,0$& 0.088(4) & 0.80(7) & 0.52(4) & 0.43(12)& 0.09(3)\\ {\em} &
$1,1,0$& -0.035(3) & 0.61(15) & 0.47(15) &0.36(34) &-0.04(3) \\ \hline
0.1275, & $0,0,0$& 0.254(8) & $-$ & 0.59(6) & $-$& $-$ \\ 0.1490 & $1,0,0$&
0.096(5) & 0.77(10) & 0.51(5) & 0.45(16)& 0.12(5)\\ {\em} & $1,1,0$&
-0.030(3) & 0.59(18) & 0.44(20) &0.30(42) &-0.03(3) \\ \hline 0.1275, &
$0,0,0$& 0.272(9) & $-$ & 0.56(7) & $-$& $-$ \\ 0.1495 & $1,0,0$& 0.105(5) &
0.76(16) & 0.52(8) & 0.51(26)& 0.17(8)\\ {\em} & $1,1,0$& -0.025(4) &
0.56(24) & 0.37(28) &0.19(57) &-0.03(4) \\ \hline 0.1325, & $0,0,0$& 0.139(5)
& $-$ & 0.66(6) & $-$& $-$ \\ 0.1485 & $1,0,0$& 0.011(3) & 0.82(7) & 0.56(5)
& 0.43(12)& 0.013(7)\\ {\em} & $1,1,0$& -0.095(2) & 0.62(16) & 0.50(16)
&0.34(31) &-0.12(9) \\ \hline 0.1325, & $0,0,0$& 0.152(5) & $-$ & 0.64(7) &
$-$& $-$ \\ 0.1490 & $1,0,0$& 0.016(3) & 0.79(10) & 0.55(6) & 0.44(16)&
0.02(1)\\ {\em} & $1,1,0$& -0.092(2)& 0.58(19) & 0.47(21) &0.29(39) &
-0.12(12) \\ \hline 0.1325, & $0,0,0$& 0.165(7) & $-$ & 0.60(8) & $-$& $-$ \\
0.1495 & $1,0,0$& 0.022(4) & 0.78(17) & 0.56(9) & 0.48(25) & 0.04(2)\\ {\em}
& $1,1,0$& -0.089(2) & 0.55(24) & 0.39(31) &0.16(53) &-0.11(16) \\ \hline
0.1375, & $0,0,0$& 0.064(3) & $-$ & 0.71(7) & $-$& $-$ \\ 0.1485 & $1,0,0$&
-0.042(1) & 0.84(8) & 0.59(5) & 0.43(10)& -0.06(2)\\ {\em} & $1,1,0$&
-0.130(6) & 0.62(17) & 0.53(16) &0.32(27) &-0.19(14) \\ \hline 0.1375, &
$0,0,0$& 0.072(4) & $-$ & 0.69(7) & $-$& $-$ \\ 0.1490 & $1,0,0$& -0.039(2) &
0.81(11) & 0.58(7) & 0.43(15)& -0.07(3)\\ {\em} & $1,1,0$& -0.129(1) &
0.57(20) & 0.50(22) &0.26(35) &-0.19(19) \\ \hline 0.1375, & $0,0,0$&
0.081(5) & $-$ & 0.64(8) & $-$& $-$ \\ 0.1495 & $1,0,0$& -0.036(2) & 0.79(17)
& 0.59(10) & 0.43(23)& -0.08(4)\\ {\em} & $1,1,0$& -0.128(1) & 0.51(24) &
0.41(32) &0.10(48) &-0.18(27) \\ \hline 0.1425, & $0,0,0$& 0.015(1) & $-$ &
0.75(8) & $-$& $-$ \\ 0.1485 & $1,0,0$& -0.068(2) & 0.87(9) & 0.62(6) &
0.41(8)& -0.13(5)\\ {\em} & $1,1,0$& -0.135(-) & 0.61(18) & 0.56(17)
&0.27(22) &-0.26(18) \\ \hline 0.1425, & $0,0,0$& 0.019(2) & $-$ & 0.72(8) &
$-$& $-$ \\ 0.1490 & $1,0,0$& -0.067(-) & 0.81(12) & 0.61(8) & 0.39(12)&
-0.15(6)\\ {\em} & $1,1,0$& -0.136(-) & 0.53(21) & 0.52(22) &0.20(28)
&-0.26(24) \\ \hline 0.1425, & $0,0,0$& 0.023(2) & $-$ & 0.67(9) & $-$& $-$
\\ 0.1495 & $1,0,0$& -0.066(1) & 0.78(18) & 0.60(11) & 0.36(19)& -0.17(10)\\
{\em} & $1,1,0$& -0.136(-) & 0.43(25) & 0.40(33) &0.02(40) &-0.21(35) \\
\hline
\end{tabular}
\caption{\it{Pseudoscalar $\rightarrow$ vector form factors
for different momenta and quark masses using the ``analytic" method. The
values reported in the table have been obtained using the ``conserved"
current.  We also give the squared momentum transfer in lattice units.}}
\label{tab:ff2}
\end{table}
\par We are interested in the form factors at  different $q^2$,
for quark masses corresponding to the physical $D$ and $K$ ($\pi$) mesons.
Thus we have to extrapolate the form factors reported in tables \ref{tab:ff1}
and \ref{tab:ff2}, both in mass and momentum.  We have proceeded as follows:
\begin{table}
\centering
\begin{tabular} {|c|c|c|c|c|c|}
\hline
{\em $K_1$} & {\em $K_2$ } & {\em $aM_{5}$ } & {\em $aM_{V}$ } & {\em
$aM_{A}$ } & {\em $aM_{S}$ }
\\ \hline
0.1275 & 0.1485 & 0.813(4) & 0.822(4) & 0.94(2) & 0.91(2) \\ 0.1275 & 0.1490
& 0.804(4) & 0.812(4) & 0.93(3) & 0.90(3) \\ 0.1275 & 0.1495 & 0.794(5) &
0.803(5) & 0.92(3) & 0.90(4) \\ 0.1275 & $K_s$ & 0.795(6) & 0.803(6) &
0.92(3) & 0.90(4) \\ 0.1275 &$K_{cr}$& 0.773(6) & 0.783(5) & 0.91(5) &
0.88(6) \\
\hline
0.1325 & 0.1485 & 0.700(4) & 0.712(4) & 0.83(2) & 0.81(3) \\ 0.1325 & 0.1490
& 0.689(4) & 0.702(4) & 0.83(3) & 0.80(3) \\ 0.1325 & 0.1495 & 0.679(5) &
0.693(5) & 0.82(4) & 0.79(5) \\ 0.1325 &$K_{s}$ & 0.680(6) & 0.693(6) &
0.82(4) & 0.79(5) \\ 0.1325 &$K_{cr}$& 0.657(6) & 0.672(6) & 0.81(5) &
0.77(7) \\
\hline
0.1375 & 0.1485 & 0.580(4) & 0.598(5) & 0.73(3) & 0.69(3) \\ 0.1375 & 0.1490
& 0.569(4) & 0.588(5) & 0.72(3) & 0.68(4) \\ 0.1375 & 0.1495 & 0.558(5) &
0.578(5) & 0.72(4) & 0.67(6) \\ 0.1375 &$K_{s}$ & 0.558(6) & 0.578(7) &
0.72(4) & 0.67(5) \\ 0.1375 &$K_{cr}$& 0.533(6) & 0.556(6) & 0.71(6) &
0.65(8) \\
\hline
0.1425 & 0.1485 & 0.450(4) & 0.479(5) & 0.63(3) & 0.56(4) \\ 0.1425 & 0.1490
& 0.438(4) & 0.468(6) & 0.62(4) & 0.55(5) \\ 0.1425 & 0.1495 & 0.425(5) &
0.457(6) & 0.62(5) & 0.54(7) \\ 0.1425 &$K_{s}$ & 0.425(6) & 0.457(8) &
0.62(5) & 0.54(7) \\ 0.1425 &$K_{cr}$& 0.396(5) & 0.433(7) & 0.60(7) &
0.51(10) \\
\hline
\end{tabular}
\caption{\it{Pseudoscalar ($M_5$), vector ($M_V$), axial ($M_A$) and
scalar ($M_S$) meson masses for mesons composed of heavy-light quarks
(dimensionless units). }}
\label{tab:om}
\end{table}

\par i) At fixed heavy quark mass and light meson momentum, $\vec p_{K,K^*}$,
the generic form factor $F$ ($F=f^+$, $A_1$ ,$...$) has been extrapolated
linearly in the light quark mass to values corresponding to the strange ($D
\rightarrow K,K^*$) or massless ($D \rightarrow \pi,\rho$) quarks.  In
ref.\cite{lubicz2}, it was shown that $SU(3)$ symmetry breaking effects are
very small, i.e. that a linear dependence of the type:
\be F= \alpha + \beta (\frac{1}{K_1} + \frac{1}{K_2}) \label{lq} \ee
describes well the behaviour of $F$ as a function of the light quark mass
($m_q^{1,2}=1/2 \, a \times ( 1/K_{1,2} - 1/K_{cr})$).  In eq.(\ref{lq})
$K_1$ and $K_2$ correspond to the masses of the final and spectator quarks
respectively.  Thus the form factors extrapolated to the strange and light
quark masses, $F^K$ and $F^{\pi}$, read $F^K= \alpha + \beta ( 1/ K_s +1
/K_{cr})$ and $F^\pi = \alpha +2 \beta /K_{cr}$, where $K_s=0.1495(1)$ is the
value of the Wilson parameter for the strange quark and $K_{cr}=0.1506(1)$
the critical one.  The value of $K_{charm}$ corresponding to the charmed
quark is $0.1379(11)$\footnote{The errors given here are slighty different
from those on ref.\cite{orsw} where the time interval $14$-$24$ was used for
the fits of the two-point functions. In the present study we find
$a^{-1}=3.6(2)$ GeV instead of $3.7(2)$ as in ref.\cite{orsw}.  For the
heavy-light mesons the interval $14-24$ is probably more appropriate, but we
have checked that the difference for the form factors is immaterial given the
statistical errors of the results.}.
\par ii) Next we have  extrapolated  $F^{K,\pi}$  in the mass of the heavy
quark according to the expression:
\be F^{K,\pi}= A+ \frac{B}{M_P} , \label{fit1} \ee
where $M_P$ is the mass of the heavy meson.  At each value of $\vec
p_{K,K^*}$, we have also extrapolated in $1/M_P$ to the D and B mesons using
the dependence expected in the heavy quark effective theory
(HQET)\cite{sliw}, see eq.(\ref{fit2}) below.  The difference of the results
obtained with the two methods are discussed later on. For $D$ decays, we have
verified that extrapolating in a different order, e.g. first in $q^2$, then in
the light quark mass, and finally in the heavy quark mass, leads to very
similar results, within the statistical errors.
\par iii)
To obtain the form factors at $q^2=0$, we have extrapolated the form factors
at $\vec p_{K,K^*}=(1,0,0)$ (which in many cases corresponds to the smallest
$q^2$, cf. tables
\ref{tab:ff1} and \ref{tab:ff2}) by assuming meson dominance:
\be F(q^2)= \frac {F(0)} {1-q^2/M^2_t} ,  \label{vmd} \ee
where $F$ can be $f^+$, $V$, $A_1$ etc. and $M_t$ is the mass of the lightest
meson exchanged in the t-channel.  Thus the vector, scalar and axial scalar
meson masses have been used for the extrapolation of $f^+$ and $V$, $f^0$ and
$A_{1,2}$ respectively.  $M_t$ is computed on the lattice, over the same
configurations, at the same heavy and light quark masses used for the
three-point functions.  In table \ref{tab:om} we report, in lattice units,
the heavy-light pseudoscalar, vector, scalar and axial masses used in the
present analysis, at different values of the Wilson parameters. In table
\ref{tab:mt} we also report the physical masses as extrapolated from the
values given in table \ref{tab:om}.  To the results in table \ref{tab:mt} we
add $M_{D_s}-M_{D_d}=(93\pm6)$ MeV and $M_{B_s}-M_{B_d}=(75\pm12)$ MeV
\footnote{In ref.\cite{orsw} were we used a time interval more apropriate  to
spectroscopy,
we found $M_{D_s}-M_{D_d}=(86\pm5)$ MeV and $M_{B_s}-M_{B_d}=(62\pm7)$ MeV.
Using a improved action\cite{apefb} in the static limit it has been found
$M_{B_s}-M_{B_d}=(76\pm10)$ MeV.  These values can be compared to the
experimental results $M_{D_s}-M_{D_d}=(99.9\pm0.7)$ MeV\cite{pdg} and
$M_{B_s}-M_{B_d}\sim(96.3\pm 4.8)$ MeV obtained from $M_{B_s}=5374.9 \pm 4.4$
MeV\cite{pipo} and $M_{B_d}=5278.6\pm2.0$ MeV\cite{pdg}.}.  In doing the
extrapolation from small $q^2$'s to $q^2=0$, the precise value of $M_t$ is
relatively unimportant.  For example we have verified that by using in all
cases the vector meson mass, the results change by about $5 \%$. \par
\begin{table}
\centering
\begin{tabular} {|c|c|c|c|c|c|}
\hline
{\em $M_{D^*}$} & {\em $M_{D^{**}}\, 1^{++}$ } & {\em $M_{D^{**}}\, 0^{++}$ }
& {\em $M_{{D_s}^*}$ } & {\em $M_{{D_s}^{**}}\, 1^{++}$ } & {\em
$M_{{D_s}^{**}}\, 0^{++}$ }
\\ \hline
$ 1.95\pm 0.01$ & $2.50 \pm 0.22$ & $2.30 \pm 0.28$ & $2.04\pm0.02$ & $2.55
\pm 0.16$ & $2.38 \pm 0.19$ \\
\hline \hline
{\em $M_{B^*}$} & {\em $M_{B^{**}}\, 1^{++}$ } & {\em $M_{B^{**}}\, 0^{++}$ }
& {\em $M_{{B_s}^*}$ } & {\em $M_{{B_s}^{**}}\, 1^{++}$ } & {\em
$M_{{B_s}^{**}}\, 0^{++}$ }
\\ \hline
 $5.27 \pm 0.01$ & $5.66 \pm 0.15$ & $5.61 \pm 0.17$ & $5.31 \pm 0.02$ &
$5.72 \pm 0.10$ & $5.68 \pm 0.11$ \\
\hline
\end{tabular}
\caption{\it{Masses in GeV predicted from the lattice for the vector,
axial and scalar excitations of the $D$ and $B$ mesons (remember that the
pseudoscalar masses $M_D$ and $M_B$ are used as an input).  These masses have
been used in several occasions, as mentioned in the text, to extrapolate the
form factors at $q^2=0$.}}
\label{tab:mt}
\end{table}
In order to show the $q^2$ dependence of the different form factors and to
compare it with the meson dominance hypothesis, we display in figs.\ref{md}
the various form factors. They are shown as a function of the dimensionless
variable $q^2/M_t^2$, for $K_H=0.1375$ and $K_L=0.1495$, which corresponds to
the meson masses closest to the physical ones for $D \rightarrow K,K^*$
decays.  For any given form factor, $M_t$ is the lattice meson mass
appropriate for that particular channel.  In the figures we have also
included the points corresponding to a final meson with momenta $(1,1,0)$ and
$(2,0,0)$. These points are reported for completeness but, given the
extremely large systematic effects, they have never been used in the
analysis.  These figures show that the behaviour is compatible with meson
dominance, at least in the range of masses and $q^2$ explored in our
simulation. It is unknown whether the meson dominance will remain valid when
we extrapolate to B decays, since the range of $q^2$ extends much further
away from the pole in that case. Notice that in ref.\cite{bbd}, using QCD sum
rules, it was found that the axial form factors do not follow the behaviour
expected on the basis of the meson dominance.
\begin{figure}[t]   
    \begin{center} \setlength{\unitlength}{1truecm} \begin{picture}(6.0,6.0)
\put(-6.0,-9.0){\special{md.ps}}

       \end{picture} \end{center} 	\vskip 2.6cm \caption[]{\it{We show
an example of the $q^2$ behaviour of the form factors. We have chosen for the
light and heavy Wilson parameters K=0.1495 and 0.1375 respectively. In each
plot the two points furthest to the left (corresponding to the momentum
assignment (1,1,1) and (2,0,0)) are only shown for the sake of illustration.
We have never used them in the fits due to their large systematic and
statistical errors. The curves correspond to the nearest pole dominance
approximation with the mass $M_t$ of the relevant meson taken from table
\ref{tab:om} and the numerator taken to fit the point closest to $q^2=0$ i.e.
the momentum assignment (1,0,0). }} \protect\label{md}
\end{figure}
\subsection{Main Systematic Effects}
 \par Before we proceed to the physics of B and D meson decays let us
summarize and discuss the main sources of systematic effects, besides
``quenching", present in our calculation:
\par a) renormalization constants: the perturbative calculation is not
unique in the sense that one has freedom in choosing the appropriate
expansion parameter. It has been suggested that the bare coupling constant
$g_0^2=6/\beta$ is not very suitable and that it should be replaced by an
effective coupling $ g^2_{eff}$
\cite{parisi}-\cite{lm2}. In our case, using the recipe of ref.\cite{lm1,lm2},
 $g^2_{eff}/g_0^2$ turns out to be $\sim 1.59$, which leads, using the
formulae of refs.\cite{mz}-\cite{smit}, to $Z_V=0.75$ and $Z_A=0.84$. The
reliability of these estimates can be tested by the non-perturbative ratios
(see also ref.\cite{mm,msv}):
\be Z_V = \frac {< \alpha \vert V_\mu^C \vert \beta >}{< \alpha \vert V_\mu^L
\vert \beta >} ,\label{ezv}\ee
where $V^{L,C}$ are the local and conserved currents respectively.  In our
simulation, we found values of $Z_V$ which vary between $0.66$ and $0.95$
depending on the matrix element, see table \ref{tab:zv}.
\begin{table}
\centering
\begin{tabular}  {|c|c|c|c|c|c|} \hline
$\vec p $ &$<V_1 V_1>$& $<K J_0 D>$& $<K J_1 D>$& $<K_3^* J_1 D>$ \\ \hline
$(0,0,0)$& $0.659(3)$&$0.87(-)$& $-$ & $-$
\\ \hline
$(1,0,0)$&$0.652(3)$&$0.94(1)$&$0.75(1)$&$0.66(2)$ \\ \hline
$(1,1,0)$&$0.645(5)$&$0.95(3)$&$0.79(3)$ & $0.69(5)$ \\ \hline
\end{tabular}
\caption{\it{Values of $Z_V$ calculated
by taking the ratio of three-point correlation functions with inserted the
``conserved" weak vector current divided by the corresponding correlation
function with the local current inserted, cf. eq.(10). The Wilson parameters
used for the three-point functions are $K_H=0.1375$ and $K_L=0.1495$.  The
results are shown for several momenta of the $K,K^*$, and for different
Lorentz components of the weak current, $J_i$,	 and $K^*$ source, $K^*_i$.
We also give $Z_V$ as derived from the light-light two-point function in the
column labelled as $<V_1 V_1>$, computed with $K_L = 0.1495$.}}
\label{tab:zv}
\end{table}
This range of values reflects uncertainties which have been interpreted as
$O(a)$ effects\cite{hmprs} and shown to be much smaller with the use of an
``improved" action \`a la Symanzik\cite{msv}.  In all our estimates we have
used the perturbative values, $Z_V=0.84$ and $Z_A=0.88$.  With the present
systematic uncertainties this choice is appropriate. The reader can easily
replace our numbers by those corresponding to his preferred value for
$g^2_{eff}$.

\par b) $O(a)$ effects: from the above discussion we expect a systematic error
of the order of $10$-$20 \%$ coming from $O(a)$ effects in the current matrix
elements.  The Fermilab group has suggested that this systematic error may be
reduced by multiplying the propagators of the heavy quark $Q$ by $\exp
(m_Qa)$ and redefining its mass\cite{kronfeld}.  This prescription, which
does not change the results for $m_Qa\ll 1$, is motivated by the tree-level
behaviour of lattice propagators.  It is expected to work for small momenta,
$ p_{\mu} a \ll 1$, when the quark mass is large, $m_Q a \sim 1$.  The
difference between the results obtained using the standard procedure (i.e.
multiplying the lattice quark propagators only by $2 K$), and those obtained
following the modified prescription at least represent some measure of the
uncertainty due to discretisation errors.
\par We observe  that the modification proposed by the
Fermilab group leads to a universal change of all the matrix elements of a
given current. Consequently, it cannot repair the fact that one finds very
different values of $Z_V$ for different matrix elements as shown in table
\ref{tab:zv}. Furthermore we have checked that $Z_V$ from $<V_1V_1>$ is
almost independent of the quark mass. Indeed $Z_V$ from $<V_1V_1>$ for
heavy-light mesons ranges from $0.643(2)$ to $0.653(3)$, which is almost
identical to the light-light case reported in table \ref{tab:zv}. We conclude
that in this case the dependence of $Z_V$ on the quark mass does not follow
the behaviour predicted in ref.\cite{kronfeld}
\footnote{This  indicates that the vacuum to vector matrix element has a
dependence on the quark mass different from that observed for the forward
matrix element of the local current\cite{kronfeld}.  Similarly one may worry
about the use of the recipe
of ref.\cite{kronfeld} for the pseudoscalar meson decay constants.}.  For
this reason we have not followed the suggestion of ref.\cite{kronfeld}. We
have instead decided to report here the form factors obtained from the local
and ``conserved" vector current and the local axial current. At the end, we
will average the results obtained from the local and conserved vector current
and add (in quadrature) as a systematic error ($\sim 10-20\%$ for $V(0)$)
their difference.  In the case of the axial current we only have results from
the local axial current, for which we also expect a $10$-$20\%$ systematic
uncertainty coming from $O(a)$ effects.
\par c) extrapolation to the physical quark masses: the extrapolation
in the light quark mass is quite natural and unlikely to be a source of an
important uncertainty within our statistical accuracy.  More delicate can be
the extrapolation in the heavy quark mass.  There are arguments, based on
HQET which allow the expansion of the form factors at fixed $\vec p_{K,K^*}$,
with $\vert \vec p_{K,K^*} \vert \ll M_P$, in inverse powers of the heavy
meson mass $M_P$\cite{sliw}.  All the relevant formulae are given in
sec.\ref{sec:slb}, where we discuss the extrapolation to the B meson.  Here
we simply state that HQET suggests that $f^+$ scales, at fixed $\vec
p_{K,K^*}$, as:
\be  f^+ =   M_P^{1/2} \, \gamma_+ \times\Bigl( 1 + \frac{\delta_+}
{ M_P} \Bigr) \label{fit2} \ee and similarly for $f^-$, $V$ and $A_{2}$,
while $A_1$ scales like $M_P^{-1/2}$.  For D mesons, the results, obtained by
extrapolating the values given in tables \ref{tab:ff1} and \ref{tab:ff2} by
using eq.(\ref{fit1}) (fit ``a") or eq.(\ref{fit2}) ( fit ``b") are reported
in table \ref{tab:semifinal} \footnote{ Below we will give the same kind of
table also for the B meson.}.  Although we have found that the $\chi^2$ is
slightly better if we fit the form factors according to eq.(\ref{fit2}), we
do not have sufficiently good data to distinguish between the two behaviours.
For completeness we also give the results obtained by using the local vector
current (in parenthesis). From this table one sees that for D mesons the
difference between the extrapolation ``a" and ``b" is rather small, $ \le
3\%$. There is more difference in the case of the vector current between the
results obtained using the conserved or the local current, in particular in
the case of $D \rightarrow K^*$. As mentioned before we include this
difference in the final error. For the axial current we are not able to
estimate the error due to the determination of $Z_A$, since we do not have
the equivalent of the conserved vector current. Besides the statistical
errors reported in the tables, we then expect for the axial form factors a
further error of order $10 - 20\%$.
\begin{table}
\centering
\begin{tabular}{|c|c|c|c|c|}
\hline
& $f^+_K(0)$ & $V(0)$ & $A_1(0)$&$ A_2(0)$\\ \hline ``a"-analytic&$0.73(72)
\pm 0.16(16)$&$0.85(1.10) \pm 0.24(30)$& $0.63 \pm 0.14$& $0.46 \pm 0.33$ \\
\hline ``b"-analytic&$0.74(72) \pm 0.16(16)$&$0.85(1.10) \pm 0.24(30)$& $0.63
\pm 0.14$& $0.46 \pm 0.33$ \\ \hline ``a"-ratio&$0.60(59) \pm
0.12(12)$&$0.84(1.09) \pm 0.24(30)$& $0.63 \pm 0.14$&$0.44 \pm 0.27$ \\
\hline ``b"-ratio&$0.61(60) \pm 0.12(12)$&$0.85(1.10) \pm 0.23(29)$& $0.63
\pm 0.14$& $0.44 \pm 0.27$ \\ \hline
\end{tabular}
\caption{\it{Semi-leptonic form factors  at zero momentum transfer
for $D \rightarrow K$ and $K^*$ using different extrapolations to the D
mesons: eq.(8), labelled as ``a", and eq.(11), labelled as ``b". The results
are reported using the method called ``analytic" or ``ratio" to extract the
form factors. The number obtained from the local vector current are reported
in parenthesis.}}
\label{tab:semifinal}
\end{table}
\section{Physics Results}
\subsection{D  Meson Decays}
\par  Given the discussion of the previous section, our best estimates
for the form factors and partial widths are those reported in the abstract
and in tables \ref{tab:final} and \ref{tab:largeurs}.  In this table we
report our results together with other calculations and experimental
determinations of the form factors.  With respect to other calculations at a
lower value of $\beta$, for example refs.\cite{victor}-\cite{lubicz2} the
errors are larger.  This is expected because the statistics is lower and
$\beta$ is higher.  It is however reassuring that, at a value of $a^{-1}$
almost twice that of refs.\cite{victor}-\cite{lubicz2}, we find compatible
results.  Moreover experiments and lattice calculations are in good agreement
for $f^+_K$, $V$ and $A_1$.  On the other hand for $A_2$ the situation is
unclear. E691 and ref.\cite{lubicz2} suggest a smaller value of $A_2$ than
E653, ref.\cite{bes} and the present work, even though the errors on this
quantity are so large that all the results are compatible. Thus for $A_2$ it
is very important to reduce both the experimental and theoretical errors.
The agreement with the other form factors is already remarkable, given the
fact that the lattice calculations have no free parameter.
\begin{table}
\centering
\begin{tabular}{|c|c|c|c|c|}
\hline
Ref.& $f^+_K(0)$ & $V(0)$ & $A_1(0)$&$ A_2(0)$\\ \hline this work &$0.65 \pm
0.18$&$0.95 \pm 0.34$&$0.63 \pm 0.14$& $0.45 \pm 0.33$ \\ \hline
\cite{victor}-\cite{lubicz2}&$0.63 \pm 0.08$&$0.86 \pm 0.10$&$0.53 \pm 0.03$&
$0.19 \pm 0.21$ \\ \hline
\cite{bes}-\cite{bes2}&$0.90\pm0.08\pm0.21$&$1.43\pm 0.45\pm0.49$&$0.83\pm
0.14\pm0.28$&
$0.59 \pm 0.14\pm0.24$ \\ \hline
\cite{wsb}& $0.76$&$1.23$&$0.88$&$1.15$ \\ \hline
\cite{wisg}& $0.76-0.82$&$1.1$&$0.8$&$0.8$ \\ \hline
\cite{bbd}& $0.6^{+0.15}_{-0.10}$&$1.1\pm 0.25$&$0.5\pm0.15$&$0.6\pm0.15$ \\
\hline
Exp. \cite{e691}&$0.70 \pm 0.08$&$0.9 \pm 0.3 \pm 0.1$&$ 0.46 \pm 0.05\pm
0.05$&$0.0 \pm 0.2 \pm 0.1$\\ \hline \hline Ref.& $A_1(q^2_{max})$ &
$V(0)/A_1(0)$ & $A_2(0)/A_1(0)$&$f^0(q^2_{max})$\\ \hline this work &$0.62
\pm 0.09 $&$1.50\pm 0.28$&$0.7 \pm 0.4 $&$0.93 \pm 0.13$ \\ \hline
\cite{lubicz2} &$0.77 \pm 0.20$ &$1.6\pm 0.2 $& $0.4 \pm 0.4$& $-$\\ \hline
\cite{bes2}&$1.27 \pm 0.16\pm0.31$&$1.99 \pm 0.22 \pm 0.33$&$0.7 \pm 0.16\pm
0.17$&$-$\\
 \hline
\cite{wsb}& $-$&$1.1$&$1.3$&$1.15$ \\ \hline
\cite{wisg}& $-$&$1.4$&$1.0$&$-$ \\ \hline
\cite{bbd}& $-$&$2.2\pm0.2$&$1.2\pm0.2$&$-$ \\ \hline
Exp. \cite{e691}& $0.54 \pm 0.06\pm 0.06$& $-$ & $-$& $-$ \\ \hline Exp.
\cite{e653}& $-$&$2.00 \pm 0.33\pm 0.16 $&$ 0.82\pm 0.23\pm 0.11$& $-$ \\
\hline \hline
\end{tabular}
\caption{\it{Semi-leptonic form factors for $D \rightarrow
K$ and $K^*$. For $f_K^+$ and $V$ we have averaged the results obtained from
the conserved and local currents and considered the difference as an error to
add in quadrature to the statistical one.}}
\label{tab:final}
\end{table}

\begin{table}
\centering
\begin{tabular}{|c|c|c|c|}
\hline
Ref.& $\Gamma(D\rightarrow K)/ 10^{10}s^{-1}$ & $\Gamma(D\rightarrow K^*)/
10^{10}s^{-1}$ & $\Gamma(D\rightarrow \pi)/ 10^{10}s^{-1}$\\ \hline this work
&$6.8\pm 3.4$&$6.0\pm 2.2 $&$0.56\pm 0.36 $\\ \hline
\cite{lubicz2}& $5.8\pm 0.15$&$5.0\pm0.9$&$0.5\pm0.2$ \\ \hline
exp. \cite{stone,ball} &$7.0 \pm 0.8$&$4.0\pm 0.7$&$0.9^{+0.5}_{-0.3}$ \\
\hline \hline Ref.&$ \Gamma(D\rightarrow \rho)/ 10^{10}s^{-1}$&
$\Gamma(D\rightarrow K^*)/\Gamma(D\rightarrow K)$ & $\Gamma_L/\Gamma_T$ \\
\hline this work &$0.50\pm 0.23$& $0.92 \pm 0.55$&$1.27\pm 0.29$ \\ \hline
\cite{lubicz2}& $0.4\pm 0.09$&$0.86\pm0.22$&$1.51\pm0.27$ \\ \hline
\cite{wsb}& $-$ & $1.14$&$0.89$ \\ \hline
\cite{wisg}& $-$ & $1.45$&$1.11$ \\ \hline
\cite{bbd}& $-$ & $0.5\pm0.15$&$0.86\pm0.06$ \\ \hline
\cite{pbu}&$-$& $0.63\pm0.09$&$-$\\ \hline
exp. \cite{stone} &$-$&$0.57\pm 0.08$&$1.15 \pm 0.17$ \\ \hline \hline

\end{tabular}
\caption{\it{Semi-leptonic partial widths for $D \rightarrow K$, $K^*$,
$\pi$ and $\rho$, using $V_{cs}=0.975$ and $V_{cd}=0.222$.  We also report
the ratio of the longitudinal to transverse polarisation partial widths for
$D \rightarrow K^*$.}}
\label{tab:largeurs}
\end{table}

\subsection{Extrapolation to B meson decays}
\label{sec:slb}
With the present lattice spacing we are unable to study directly the B mesons
in numerical simulations. However indirect information on B physics may be
available using the following strategy.  One first studies the scaling
behaviour of a given physical quantity in the region of the charm quark mass.
Then one computes the same quantity in the effective (``static") theory, i.e.
in the limit in which the heavy quark mass is infinite.  By interpolating
between the two results with the help of the expected scaling laws, one can
predict that physical quantity for the B meson. The value in the static limit
reduces the uncertainty due to the extrapolation from the charm region. This
strategy has proved to be effective for the pseudoscalar decay constant, see
for example refs.\cite{orsw,wupp}.
\par At our large value of $\beta$, the small statistics is not sufficient to
compute semi-leptonic form factors in the static limit.
Nevertheless we can study the scaling behaviour of the form factors and try
an extrapolation to the $b$ quark. With the actual errors, our predictions
remain necessarily at a semi-quantitative level. The study is however
interesting in itself. We have found that the expected dependence of the form
factors in $1/M_P$, see eq.(\ref{fit2}), is compatible with our results.  For
some of the form factors however the corrections to the static results in the
charm region may be sizeable.  In the following we explain the extrapolation
of the form factors, computed at several values of the heavy quark mass, to
the B meson.\par On the basis of HQET, up to $O(1/M_P^2)$, up to logarithmic
corrections, one expects the following behaviour for the relevant form
factors \cite{sliw}:
\bea \frac{f^+}{M_P^{1/2}}&=& \gamma_+ \times \Bigl( 1+ \frac{\delta_+}{M_P}
\Bigr) \,\,\,\,\,\,\,\,\,\,\,\,\,
\frac{V}{M_P^{1/2}} = \gamma_V \times \Bigl( 1+ \frac{\delta_V}{M_P}
\Bigr) \nonumber \\ \frac{A_2}{M_P^{1/2}}&=& \gamma_2 \times \Bigl( 1+
\frac{\delta_2}{M_P} \Bigr)
\,\,\,\,\,\,\,\,\,\,\,\,\,
A_1 M_P^{1/2}= \gamma_1 \times \Bigl( 1+ \frac{\delta_1}{M_P}
\Bigr) \label{scale} \eea
The expansions given in eqs.(\ref{scale}) become valid in the limit of large
$m_Q$, at fixed momentum $\vec p$ of the light meson (in the frame where the
heavy meson is at rest) and when $\vert \vec p \vert \ll m_Q \sim M_P$. The
above conditions are always satisfied for $q_{max}$, when the initial and
final mesons are both at rest. Provided we are not too close to the meson
pole in the t-channel, the dependence on the light mass is smooth and it is
well described by eq.(\ref{lq}).  We can then extrapolate in the heavy quark
mass according to eqs.(\ref{scale}). The points corresponding to $\vec
p=(1,0,0)$ also satisfy the above conditions. However, in the range of masses
where we compute the form factors, $m_Q \sim m_{charm}$, these points
correspond to $q^2 \sim 0$. This is due to the fact that the heavy meson
masses are not so heavy in comparison with $\vert \vec p \vert$ and that the
light mesons are not really light enough.  Thus, while the typical
$\sqrt{q^2}$ on our lattice is at most $1.4$ GeV, the extrapolation in the
heavy quark mass will bring us to $\sqrt{q^2}=4.2$ GeV ($q^2=M_B^2-2 * M_B
\sqrt{M_\pi^2 +(2 \, \pi /24 \, a)^2}+M_\pi^2$), and similarly for $B
\rightarrow \rho$.  The uncertainty involved in such an extreme extrapolation
must be borne in mind.  The validity of the extrapolation is partially
justified only because the range in $1/M_P$ is relatively small and most of
the form factors have a smooth behaviour.  The case $(1,1,0)$ is more
complicated because, with the heavy quark masses at hand, we have $\vert \vec
p \vert \sim M_P$ where the expansion of ref.\cite{sliw} is questionable.
Anyhow the errors are so large that we have not used these points. \par
\begin{figure}[t]   
    \begin{center} \setlength{\unitlength}{1truecm} \begin{picture}(6.0,6.0)
\put(-6.0,-9.0){\special{scala.ps}}

       \end{picture} \end{center}
\vskip 2.6cm
    \caption[]{\it{Form factors extrapolated to the chiral limit for the
light quark, as a function of the inverse pseudoscalar mass ($1/M_P$) for the
momentum assignment (1,0,0).  The crosses are the lattice points, the
diamonds are the extrapolation to the D and B meson. Notice that the D is
very close to a lattice point.  The points corresponding to the lightest
heavy quark mass (furthest to the right) has not been used in the fits. }}
\protect\label{scaling}
\end{figure}

In fig.\ref{scaling} we give the form factors, at the value of $q^2$
corresponding to the momentum assignment $(1,0,0)$, extrapolated to the
chiral limit in the light quark mass, as a function of $1/M_P$ (crosses). In
the figure we also give the value extrapolated to the $D$ and $B$ mesons
(diamonds).  The figure shows that the points are smoothly extrapolated, even
in the case of $V$ since $\gamma_V$ is small. This is encouraging for further
studies with higher statistics.  In table \ref{tab:scale} we report the
values of $\gamma_{+,V,1,2}$ and $\delta_{+,V,1,2}$ for $\vec p =(0,0,0)$ and
$(1,0,0)$, in physical units. We tried also $(1,1,0)$, but the errors turned
out be of order $100 \%$ and we have not reported them in the table.  We
notice that the first correction in $1/M_P$ is small in the case $A_1$.  This
is also the case for $f^+$, at $(1,0,0)$, even though, because of the large
error, we cannot exclude a slope $\sim 0.7$ GeV. The same can be said for
$A_2$, whose value anyhow is badly determined even before the extrapolation.
In the case of $V$ we find instead a rather large correction. We believe that
this is an interesting and rather unexpected result, which deserves an effort
to reduce the size of the statistical error. Taking our present errors into
account we cannot exclude that the large value of $\delta_V$ is correlated to
the small value of $\gamma_V$ and that it will disappear with more accurate
results.
\begin{table}
\centering
\begin{tabular}{|c|c|c|c|c|}
\hline
$\vec p$ &$\gamma_+ \, \rm{GeV}^{-1/2}$ & $\gamma_V \, \rm{GeV}^{-1/2}$ &
$\gamma_1 \,\rm{GeV}^{+1/2}$& $\gamma_2 \, \rm{GeV}^{-1/2}$\\ \hline
$(0,0,0)$ &$-$&$-$&$0.96 \pm 0.16$& $-$ \\ \hline $(1,0,0)$ &$0.39 \pm
0.25$&$0.29 \pm 0.12$&$1.05 \pm 0.25$& $0.44 \pm 0.25$ \\ \hline \hline $\vec
p$& $\delta_+$ GeV & $\delta_V$ GeV & $\delta_1$ GeV &$\delta_2$ GeV\\ \hline
$(0,0,0)$ &$-$&$-$&$-0.33 \pm 0.09$& $-$ \\ \hline $(1,0,0)$ &$0.0 \pm
1.1$&$1.9 \pm 1.3$&$-0.46 \pm 0.22$& $-0.6 \pm 0.8$ \\ \hline
\end{tabular}
\caption{\it{The coefficients of the $1/m_Q$ expansion of the form
factors defined in eqs.(12).}}
\label{tab:scale}
\end{table}
\par {}From the numbers given in the table we can predict the form factors of
the
B mesons. We give all the form factors at $q^2=0$. They have been obtained
from the form factors at $\vec p =(1,0,0)$, by using the meson dominance,
with the mass of the mesons exchanged in the t-channel reported in table
\ref{tab:mt}.  These masses have been obtained by fitting the mass difference
$\Delta M=M_{P^*}-M_P$ (vector case) and $\Delta M=M_{P^{**}}-M_P$ (axial
case) as $\Delta M=A_M + B_M/M_P$. The results are reported in table \ref
{tab:mt}. They are rather close to the experimental masses when known, and
also to the masses computed in ref.\cite{wsb}, i.e.  $M_{B^*}=5.32$ GeV and
$M_{B^{**},1^{++}}=5.71$ GeV, except for the scalar mass which in
ref.\cite{wsb} is very large: $M_{B^{**},0^{++}}=5.99$ GeV.  The differences
between the extrapolations using different sets of masses turn out to be
small and may be smaller than the error induced by the assumption of the
meson dominance.  The results are reported in table \ref{tab:extrab}, label
``b", together with the results of refs.\cite{wsb}-\cite{ball} for
comparison.  In the evaluation of the errors we have also taken into account
of the difference between the results obtained using the local or the
conserved current. To show the stability of the results with respect to a
different extrapolation, we have also reported the values obtained with the
naive scaling given in eq.(\ref{fit1}), labelled as ``a". \par In the D meson
the difference between the HQET scaling laws and naive scaling was
immaterial. The differences remain small in the B case and with our errors we
cannot distinguish the two behaviours.  Notice that, because of the different
scaling laws, $A_2/A_1$ increases with the mass of the heavy quark and can be
$\sim 2$ for $B \rightarrow
\rho$.
\par For most of the form factors the predictions of ref.\cite{wisg}
are much lower than all the others, cf. table \ref{tab:extrab}.  These result
in a much larger estimate of $\vert V_{ub}\vert$, for a given experimental
branching ratio. Notice however that in ref.\cite{wisg} the form factor is
computed at $q^2_{max}$ and then a ``tempered'' exponential dependence on
$q^2$ is assumed. This $q^2$ yields a dramatic suppression at small $q^2$ for
B meson decays, where the range in $q^2$ is very large.

\begin{table}
\centering
\begin{tabular}{|c|c|c|c|c|}
\hline
Ref.& $f^+(0)$ & $V(0)$ & $A_1(0)$&$ A_2(0)$\\ \hline this work``a" &$0.28
\pm 0.14$&$0.37 \pm 0.14$&$0.24 \pm 0.06$& $0.39 \pm 0.24$ \\ \hline

this work ``b" &$0.33 \pm 0.17$&$0.40 \pm 0.16$&$0.21 \pm 0.05$& $0.47 \pm
0.28$ \\ \hline
\cite{wsb}& $0.33$&$0.33$&$0.28$&$0.28$ \\ \hline
 \cite{wisg} &$0.09$&$0.27$&$0.05 $& $0.02$ \\ \hline
\cite{ball} &$0.26 \pm 0.02$&$0.6\pm 0.2$&$0.5 \pm 0.1$&
$0.4 \pm 0.2$ \\ \hline \hline Ref.& $-$ & $V(0)/A_1(0)$ &
$A_2(0)/A_1(0)$&$-$\\ \hline this work ``a" &$-$& $1.4 \pm 0.5$&$1.5\pm 0.8$&
$-$ \\ \hline this work ``b" &$-$& $1.7 \pm 0.6$&$2.3
\pm 1.1$ &$-$
 \\ \hline
\cite{wsb}& $-$ & $1.0$&$1.0 $ &$-$
 \\ \hline \hline
\end{tabular}
\caption{\it{Semi-leptonic form factors for $B \rightarrow
\pi$ and $\rho$. For $f^+$ and $V$ we have averaged the results obtained
from the conserved and local currents and considered the difference as an
error to add in quadrature to the statistical one. The label ``a" refers to
the naive extrapolation in $1/M_P$, eq.(8), and label ``b" to the
extrapolation given in eqs.(12) respectively. To extrapolate to zero momentum
transfer we have used the masses of table 6.}}
\label{tab:extrab}
\end{table}
\par {}From the numbers reported in table \ref{tab:extrab}, we can give a very
rough estimate
of the $B \rightarrow \pi$ and $B \rightarrow \rho$ branching ratios.  The
branching ratios are obtained from the form factors in the table, by assuming
meson dominance for their dependence on $q^2$, see eq.(\ref{vmd}).  To get an
estimate of the errors, we have allowed the form factors to vary in all the
possible ways by one $\sigma$ within the statistical errors and to vary in all
possible ways among the values obtained with different extrapolations in
$1/M_P$, fits ``a" and ``b".  In this way we can partially account for the
uncertainty coming from the extrapolation and the $q^2$ dependence of the
form factors. We finally get:
\be \Gamma(B \rightarrow \pi l \nu_l)= \vert V_{ub} \vert^2
(12 \pm 8) \times 10^{12} s^{-1}\ee and
\be \Gamma(B \rightarrow \rho l \nu_l)= \vert V_{ub} \vert^2
(13 \pm12) \times 10^{12} s^{-1} \label{brr}
\ee

Thus the errors are still too large to give more than an upper bound. Since
experiment, on its own side, only gives upper bounds for the considered
branching ratios, we cannot with our present accuracy extract any information
on $V_{ub}$. We mainly want to stress the feasibility of this analysis,
provided one disposes of larger statistics.
\section*{Acknowledgments}
\par We thank R. Sarno for her participation in the early stages
of this work. We thank A. Le Yaouanc for fruitful discussions.  C.R.A., M.C.,
V.L. and G.M. acknowledge the partial support of the MURST, Italy, and INFN.
The work of S.G., G.S. and K.S was supported by Deutsche
Forschungsgemeinschaft, grant Schi 257/3-1 3-2. C.T.S. acknowledges the
support of the Science and Engineering Research Council through the arward of
a Senior Fellowiship. We warmly thank the staffs of CINECA, PSI (Orsay), CCVR
(Polytechnique) and H\"ochstleistungsrechenzentrum J\"ulich for their
continuous and effective support.

\end{document}